\documentclass{ws06-procs}

\begin{document}

\title{Magnetic-Optical Filter}

\author{Ilaria Formicola\footnote{e-mail: ilariaformicola@libero.it}, Andrea Longobardo\footnote{e-mail: andrea.longobardo@tele2.it}, Ciro Pinto\footnote{e-mail: ciropinto1982@libero.it},\\ Pierluigi Cerulo\footnote{e-mail:
lizard031@hotmail.it}}

\address{University of Naples Federico II, Complesso Universitario di Monte Sant'Angelo, Via Cinthia, I-80126, Naples, Italy}

\maketitle

\abstracts{Magnetic-Optical Filter (MOF) is an instrument suited
for high precision spectral measurments for its peculiar
characteristics. It is employed in Astronomy and in the field of
the telecommunications (it is called FADOF there). In this brief
paper we summarize its fundamental structure and functioning.}

Magnetic-Optical Filter (MOF) was developed in the 60's in Rome by
professor Cacciani. Its more important features are good spectral
resolution, high transmission, high field of view and an absolute
spectral reference. In Naples, at the {\itshape{Osservatorio
Astronomico di Capodimonte}} (OAC), it's utilized in the VAMOS
(Velocity And Magnetic Observations of the Sun) project, whose aim
is to measure Solar surface's velocity field and magnetic field
along the line of sight. Moreover the high signal-to-noise ratio
of the MOF permits its use in the telecommunications too (here
it's called FADOF, Faraday Anomalous Dispersion Optical Filter).

In the VAMOS instrument, MOF consists of a potassium vapours cell
with a magnetic field (about 1400 G) along its optic axis,
interposed between two crossed linear polarizers. In order to
understand how this works, we have to  recall the Zeeman effect.
Let's consider the atomic transition from the level with $l = 1$
to that with $l = 0$ (where $l$ is the angular momentum quantum
number): in absence of magnetic field, there is only an emission
line. If we are in presence of a magnetic field, the degeneration
of the level with $l = 1$ is removed bringing to three different
states with three different values of the atomic quantum number
$m$ (magnetic moment quantum number) and we can see no more
{\itshape{one}} emission line, but {\itshape {three}} emission
lines characterized by different states of polarization. In fact
two of these emission lines are circularly polarized, respectively
right-handed ($\sigma^+$) and left-handed ($\sigma^-$), around the
magnetic field direction, the other ($\pi$) is linearly polarized
along the magnetic field, so, when we observe along this direction
(it's our case) we can't see this last component. MOF is based on
two effects: the Righi Effect and the Macaluso-Corbino Effect.
Righi Effect is Zeeman effect in absorption: solar light (not
polarized) arrives on the first polarizer which transforms it in
linearly polarized light (let's recall that linearly polarized
light can be viewed as half right circularly polarized and half
left circularly polarized); then the cell absorbs half of  the
light intensity at $\sigma^+$ and $\sigma^-$ wavelengths and the
second polarizer cuts half of the light intensity at $\sigma^+$
and $\sigma^-$ wavelengths and cuts totally the other wavelengths.
So, at the output, we should see only two peaks at the Zeeman
wavelengths, but the net output of the filter is characterized by
the presence of the Macaluso-Corbino Effect, too. This consists in
a rotation of the polarization plane caused by a difference in
refraction index values at the two Zeeman wavelengths in the cell.
Higher is the temperature of the cell, stronger is the
Macaluso-Corbino Effect which shows itself as two additional
symmetric peaks, the distance between which increases linearly
with temperature (in the range considered in our work). In figure
1 we report the transmission profile in which is visible the sum
of the two effects.

\begin{figure}[ht]
\centerline{\epsfxsize=4.5cm\epsfbox{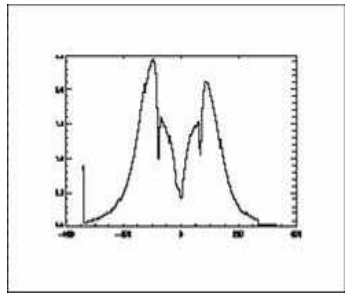}}
\caption{Transmission profile of MOF at the temperature of
160$^o$C. The Righi and Macaluso-Corbino effects are visible in
the figure; the first effect manifests itself as the two plateaux
between the two external peaks of the Macaluso Corbino Effect}
\end{figure}

To calibrate in wavelength the profiles we measured by a laser
diode system, we used the Zeeman effect. In fact, by using the
laser diode, we obtained profiles in function of the voltage
applied to the laser cavity. Because we knew the difference
$\Delta\lambda$ between the two Zeeman wavelengths (it's given by
the product $4.67 \times 10^{-13}\lambda^2 {g}\mathbf{B}$,
$\lambda$ is the unperturbed wavelength (in Angstrom), $g$ a
factor containing the Lande' factor and $\mathbf{B}$ is the
magnetic field's intensity), we could obtain profiles in function
of the wavelength. Profiles utilized in the calibration were that
obtained at  lower temperatures ($70^{\circ}C$, $80^{\circ}C$,
$90^{\circ}C$), where Zeeman lines were easier to detect.

MOF and Wing Selector (WS) form the basis of the VAMOS instrument.
WS' role is to select only one of the two MOF output lines. It's
composed by a quarter-wave plate and a cell analogous to MOF's
cell. If the plate's transmission axis forms with the optic axis
an angle of 45 degrees, light's polarization becomes right
circular and so the  cell cuts the $\sigma^+$ component, while
leaves the $\sigma^-$ one to pass; if the plate's transmission
axis forms with the optic axis an angle of -45 degrees, we have
the opposite situation and only the $\sigma^-$ component passes.

\end{document}